\documentclass[a4paper]{article}
\usepackage{mathrsfs}
\usepackage{graphicx}
\usepackage{latexsym}
\usepackage{amsmath}
\usepackage{amssymb}
\usepackage{textcomp}
\usepackage{amsbsy}
\usepackage{color}
\usepackage{epstopdf}

\title{\large \bf A possible testbed for warped extra dimension from the angle of Buchdahl's limit}
\author{Tanmoy Paul\footnote{E-mail address: pul.tnmy9@gmail.com}\\
Department of Theoretical Physics,\\
Indian Association for the Cultivation of Science,\\
2A $\&$ 2B Raja S.C. Mullick Road,\\
Kolkata - 700 032, India.\\[10mm]}
\date{}
\begin{document}
\maketitle

\begin{abstract}
We consider a five dimensional AdS warped spacetime in presence of a massive scalar field in the bulk. The scalar field potential 
fulfills the requirement of modulus stabilization even when the effect of backreaction of the stabilizing field is 
taken into account. In such a scenario, we investigate the possible role of modulus field on a compact stellar structure 
from the perspective of four dimensional effective theory. Our result reveals that in the presence of the modulus field, 
the upper bound of mass-radius ratio (generally known as Buchdahl's limit) of a star can go beyond the general 
relativity prediction. Interestingly this provides a natural testbed for the existence of such higher dimensional modulus field.
\end{abstract}
\newpage

\section{Introduction}
Ever since the original proposal of Kaluza-Klein (KK) regarding the existence of extra spatial 
dimension(s) it is often believed that our  universe is  a 3-brane embedded in a higher dimensional 
spacetime and  is described  through  a low energy 
effective theory on the brane carrying the signatures of extra dimensions \cite{kanno,shiromizu}. Depending on 
different possible compactification schemes for the extra dimensions, a large number of models 
\cite{arkani,horava,RS,kaloper,cohen,burgess,chodos} have been constructed, and 
their predictions are yet to be observed in the current experiments.\\

Among various extra dimensional models proposed over the last several decades, Randall-Sundrum (RS) warped extra dimensional model \cite{RS} earned a 
special attention since it can resolve the gauge hierarchy problem without introducing any 
intermediate scale (between Planck and TeV) in the theory. In RS model 
the interbrane separation (known as modulus or radion) is assumed to be $\sim$ Planck length and 
generates the required hierarchy between the branes.\\

A suitable potential with a stable minimum is therefore needed for modulus stabilization. Goldberger 
and Wise (GW) proposed a useful stabilization mechanism \cite{GW} by introducing 
a massive scalar field in the bulk with appropriate boundary values. Though the backreaction 
of the stabilizing scalar field was originally neglected 
in GW proposal, its  implications are subsequently studied in \cite{csaki, tp2}. It has been demonstrated in \cite{tp2} 
that the modulus of RS scenario can be stabilized using GW prescription even by incorporating the 
backreaction of the stabilizing field. Not only that, even the 
stable value of the modulus appears as a parameter in the low 
energy effective theory on the brane, but it's fluctuation about that stable value leads 
to dynamical modulus (or radion) field which couples to the fields on the 
observable brane. This attracted  
a large volume of work on phenomenological and cosmological implications \cite{csaki,GW_radion,julien,wolfe,sumanta,sorbo} 
of modulus field  in RS warped geometry model. This radion phenomenology 
along with the study of RS graviton \cite{dhr,rizzo,yong,dhr1,thomas} 
are considered to be the testing ground of warped extra dimensional models in collider experiments \cite{atlas1,atlas2}.\\

Apart from phenomenological setup, here we are interested to provide a possible testing ground for the existence of warped extra dimension 
from the angle of stable stellar structure namely the Buchdahl's limit.\\

There have been considerable interest in the compactness limit of any stellar structure, 
which, originally initiated by Buchdahl, indicates that under reasonable assumptions the minimum radius of a star 
has to be greater than $(9/8)$ of its Schwarzschild radius \cite{1,2,3}. These assumptions involve nature of the density 
of the star, which has to be decreasing outwards and also the interior solution has to be matched with an exterior one. This raises an intriguing 
question, how is the above limit modified if one considers a theory of gravity different from general relativity. This resulted 
into a large number of work quite extensively in recent times \cite{4,5,6,7,8,10,11,12,13,14} (see also \cite{15,16,17,18,19}).\\

The important questions that remain are:
\begin{enumerate}
 \item How does the compactness limit (known as Buchdahl's limit) of a stellar structure modify due to the presence 
 of radion field which carries the footprint of compactified warped extra dimension on our visible universe?
 
 \item Can the modified Buchdahl's limit be a possible testing ground for such compactified extra dimension?
\end{enumerate}

We address these questions in the present paper from the perspective of four dimensional effective theory.\\

Our paper is organized as follows. In section [2], we describe the model. In section [3] we find the possible modifications of 
Buchdahl's limit and its implications are discussed. In section [4], the interior spacetime of the stellar object is 
matched with a suitable exterior one. Finally we end the paper with some conclusive remarks.\\

\section{The model}
We consider a five dimensional AdS spacetime involving one warped and compact 
extra spacelike dimension. The spacetime is $S^1/Z_2$ orbifolded along 
the extra dimensional angular coordinate $\phi$, where the fixed points $\phi=(0,\pi)$ are 
identified with two 3-branes ($3+1$ dimensional), known as Planck (or hidden), TeV (or visible) brane 
respectively. Our usual four dimensional universe is the TeV brane and emerges as 4D effective theory. 
The opposite brane tensions along with the finely 
tuned five dimensional cosmological constant serve as energy-momentum tensor of the aforementioned configuration.\\

In higher dimensional braneworld scenario, the stabilization of extra dimensional
modulus is a crucial aspect and needs to be addressed carefully. 
It has been demonstrated by Goldberger and Wise that the modulus corresponding to the
 radius of the extra dimension in warped geometry models can be stabilized \cite{GW} by invoking a massive 
scalar field in the bulk with non zero value on the branes.\\

Keeping the stabilization mechanism in mind, the braneworld setup considered in the 
present context is represented by the following action:
\begin{eqnarray}
 S&=&\int d^5x \sqrt{-G} \bigg[M^3R - \Lambda - (1/2)G^{MN}\partial_M\Phi\partial_N\Phi-V(\Phi)\bigg]\nonumber\\
 &-&\int d^5x \sqrt{-G}\bigg[\lambda_{hid}\delta(\phi) + \lambda_{vis}\delta(\phi-\pi)\bigg]
 \label{action}
\end{eqnarray}
where $M$ is the five dimensional Planck scale, $G_{MN}$ is the 
five dimensional metric. $\Lambda$ 
symbolizes the bulk cosmological constant, $\Phi$ is the 
scalar field endowed with a potential $V(\Phi)$, 
 $\lambda_{hid}$, $\lambda_{vis}$  are the self interactions of scalar field (including brane 
tensions) on Planck, TeV branes.\\

We consider the metric ansatz as follows,
\begin{equation}
 ds^2 = \exp{[-2A(\phi)]}\eta_{\mu\nu}dx^{\mu}dx^{\nu} + r_c^2d\phi^2
 \label{metric ansatz}
\end{equation}
where $r_c$ is the compactification radius and $A(\phi)$ is termed as warp factor. For simplicity we assume that 
the bulk scalar field depends only on the extra dimensional coordinate ($\phi$). 
Thus the 5-dimensional Einstein's and scalar field equations 
for this metric can be written as,
\begin{eqnarray}
 \frac{4}{r_c^2} A'^2(\phi) - \frac{1}{r_c^2}A''(\phi)&=&-(2\kappa^2/3)V(\Phi), \nonumber\\ 
&-&(\kappa^2/3)\sum \lambda_{i}(\Phi)\delta(\phi-\phi_{i}),
 \label{equation1}
\end{eqnarray}

\begin{equation}
  \frac{1}{r_c^2}A'^2(\phi)= \frac{\kappa^2}{12r_c^2}\Phi'^2 -(\kappa^2/6)V(\Phi)
 \label{equation2}
\end{equation}

\begin{eqnarray}
 \frac{1}{r_c^2}\Phi''(\phi)=\frac{4}{r_c^2}A'\Phi' + \frac{\partial V}{\partial \Phi} 
+ \sum \frac{\partial \lambda_{i}}{\partial \Phi}\delta(\phi-\phi_{i}),
 \label{equation3}
\end{eqnarray}

where  $M^3=(1/2\kappa^2)$. Here index $i$ is used to designate the two branes and prime denotes the derivative with 
respect to $\phi$. From the above equations, 
the boundary conditions of $A(\phi)$ and $\Phi(\phi)$ are obtained as,
\begin{equation}
 \frac{1}{r_c}[A'(\phi)]_{i} = (\kappa^2/3)\lambda_{i}(\Phi_{i})
 \label{bc1}
 \end{equation}
 and
 \begin{equation}
  \frac{1}{r_c}[\Phi'(\phi)]_{i}=\partial_{\Phi}\lambda_{i}(\Phi_{i}).
 \label{bc2}
 \end{equation}
Square brackets in the above two equations represent the jump of the 
corresponding variables on the branes. In order to get an analytic 
solution, let us consider the form of the scalar field potential 
as \cite{csaki},
\begin{equation}
 V(\Phi) = (1/2)\Phi^2(u^2+4uk) - (\kappa^2/6)u^2\Phi^4
 \label{scalar potential}
\end{equation}
where $k=\sqrt{-\kappa^2\Lambda/6}$. The potential contains quadratic
 as well as quartic self interaction of the scalar field. Moreover it may be 
noticed that the mass and quartic 
coupling of the field $\Phi(\phi)$ are connected by a common free parameter $u$. 
Using this form of the potential, one obtains a 
solution of $A(\phi)$ and $\Phi(\phi)$ as follows,
\begin{eqnarray}
 A(\phi) = kr_c|\phi| + (\kappa^2/12)\Phi_{P}^2\exp{(-2ur_c|\phi|)}
 \label{solution of warp factor}
 \end{eqnarray}
 and
 \begin{eqnarray}
 \Phi(\phi) = \Phi_{P}\exp{(-ur_c|\phi|)},
 \label{solution of scalar field}
\end{eqnarray}

where $\Phi_{P}$ is taken as the value of the scalar field on the Planck brane.
 Using these solutions, $\lambda_{hid}$ and $\lambda_{vis}$ can be obtained from the 
boundary conditions (eqn.(\ref{bc1}) and eqn.(\ref{bc2})) as,
\begin{eqnarray}
 \lambda_{hid} = 6k/\kappa^2 - u\Phi_{P}^2
 \label{planck brane tension}
 \end{eqnarray}
 and
 \begin{eqnarray}
 \lambda_{vis} = -6k/\kappa^2 + u\Phi_{P}^2\exp{(-2u\pi r_c)}.
 \label{planck brane tension}
 \end{eqnarray} 
 
 In order to introduce the radion field, we consider a fluctuation of the inter-brane separation 
 around the stable configuration $r_c$ \cite{GW_radion}. This fluctuation can be treated as a field $T(x)$ (known as radion field) 
 and for simplicity this new field is assumed to be the function of brane coordinates only. Then the metric takes 
 the following form \cite{GW_radion}:
 \begin{eqnarray}
  ds^2 = \exp{[-2A(x,\phi)]}g_{\mu\nu}(x)dx^{\mu}dx^{\nu} + T^2(x)d\phi^2,
   \label{metric ansatz 2}
 \end{eqnarray}
 
 where $g_{\mu\nu}(x)$ is the induced on-brane metric and $A(x,\phi)$ has the following form,
 \begin{eqnarray}
  A(x,\phi) = k\phi T(x) + \frac{\kappa^2\Phi_{P}^2}{12}\exp{[-2u\phi T(x)]}.
   \label{warp factor 2}
 \end{eqnarray}
 
 Consequently $\Phi(x,\phi)$ can be obtained from eqn.(\ref{solution of scalar field}) by replacing 
 $r_c$ by $T(x)$ i.e.
 \begin{eqnarray}
  \Phi(x,\phi) = \Phi_{P}\exp{(-uT(x)\phi)}
 \label{solution of scalar field 2}
 \end{eqnarray}
 
 Plugging back the solutions presented in eqns. (\ref{warp factor 2}), (\ref{solution of scalar field 2})
 into original five dimensional action (in eqn. (\ref{action})) and integrating over $\phi$ yields 
the four dimensional effective action as follows
\begin{eqnarray}
 S_{eff} = \int d^4x \sqrt{-g} \bigg[\frac{M^3}{k}R_{(4)} - \frac{1}{2}g^{\mu\nu}\partial_{\mu}\Psi\partial_{\nu}\Psi 
 - U_{rad}(\Psi)\bigg]
 \label{effective action}
\end{eqnarray}
where $R_{(4)}$ is the Ricci scalar formed by $g_{\mu\nu}(x)$. Moreover,  
$\Psi(x) = \sqrt{\frac{24M^3}{k}} e^{-A(x,\pi)}$ 
(with $A(x,\pi)$ given in eqn.(\ref{warp factor 2})) is the 
canonical radion field and $U_{rad}(\Psi)$ is the radion potential with the following form \cite{tp2}
\begin{eqnarray}
 U_{rad}(\Psi) = u^2\Phi_P^2 \bigg[\frac{1}{u}\bigg(1 - e^{(2u+4k)\pi T(\Psi)}\bigg) 
 - \frac{\kappa^2\Phi_P^2}{8(u+k)}\bigg(1 - e^{(4u+4k)\pi T(\Psi)}\bigg)\bigg],
 \label{radion potential}
\end{eqnarray}
where $\kappa\Phi_P$ ($=\frac{\Phi_P}{M^{3/2}}$) is taken to be less than unity in order to ensure 
to validity of the classical solution. Moreover, $T(\Psi)$ is given by the expression: 
$\Psi(x) = \sqrt{\frac{24M^3}{k}} \exp{\big[-k\pi T(x) - \frac{\kappa^2\Phi_P^2}{12}e^{-2u\pi T(x)}\big]}$, as defined 
earlier. Using this relation between $T(x)$ and $\Psi(x)$, we obtain the minimum of the radion potential $U_{rad}(\Psi)$ at,
\begin{eqnarray}
 <\Psi> = \sqrt{\frac{24M^3}{k}} \bigg(\frac{2\sqrt{1+\frac{2k}{u}}}{\kappa\Phi_P}\bigg)^{k/u} 
 \exp{\bigg[-\frac{1}{3}\bigg(1+\frac{2k}{u}}\bigg)\bigg],
 \nonumber
\end{eqnarray}
 which immediately leads to the stabilized value of the modulus as \cite{tp2},
\begin{eqnarray}
  k\pi <T(x)>&=&k\pi r_c\nonumber\\ 
  &=&\frac{k}{u}\ln{\{\frac{\kappa\Phi_{P}}{2\sqrt{1+\frac{2k}{u}}}\}}.
   \label{stabilized modulus}
\end{eqnarray}

However our entire analysis of finding the stabilization condition in eqn.(\ref{stabilized modulus}) 
is valid only for $u>0$. In this context one can easily 
check that the radion potential produces no minima for $u<0$. Hence the parameter 
  $u$ is confined in positive regime in order to make a stable configuration for this braneworld scenario.\\
  
 On projecting the bulk gravity on the brane, the extra degrees of freedom of $R^{(5)}$ (with respect 
 to $R_4$) appears as a scalar field (the radion field), symbolized by $\Psi(x)$ in the four dimensional 
 effective action (see eqn.(\ref{effective action})). For such on-brane theory, we are interested to explore 
 the effect of radion field on stellar structure. Thus we further consider an extra matter density 
 ($L_{mat}$, confined on the brane) which acts as the ingredients of the star. Taking $L_{mat}$ into 
 account, the final form of 4D effective action is as follows,
  \begin{eqnarray}
   A_{eff}&=&S_{eff} + \int d^4x \sqrt{-g} L_{mat}\nonumber\\
   &=&\int d^4x \sqrt{-g} \bigg[\frac{M^3}{k}R_{(4)} - \frac{1}{2}g^{\mu\nu}\partial_{\mu}\Psi\partial_{\nu}\Psi 
 - U_{rad}(\Psi) + L_{mat}\bigg].
 \label{final effective action}
  \end{eqnarray}
  
  Therefore the radion field $\Psi$ (originated from extra dimension) and 
  $L_{mat}$ serve as energy-momentum tensor in the 
  four dimensional effective action. As the stellar interior is concerned, $L_{mat}$ 
  is taken to be a perfect fluid with energy-momentum tensor given by 
  $T^{\mu}_{\nu}$(matter) $= diag\big(-\rho, p, p, p\big)$. Moreover the radion field 
  contributes as,
  \begin{eqnarray}
   T_{\mu\nu}(\Psi) = \partial_{\mu}\Psi\partial_{\nu}\Psi 
   - g_{\mu\nu}\bigg[\frac{1}{2}g^{\mu\nu}\partial_{\mu}\Psi\partial_{\nu}\Psi + U_{rad}(\Psi)\bigg].
   \label{em for radion}
  \end{eqnarray}
  
  This completes our preliminary discussion and provides the necessary steps 
  that we will require in the next section while discussing the effect of radion field on a stellar structure 
  from the perspective of effective four dimensional theory (described by eqn. (\ref{final effective action})).\\
  
  \section{Buchdahl's limit on stellar structure in presence of radion field}
  As mentioned earlier, we want to investigate the possible modifications of Buchdahl's limit 
  due to the presence of radion field, thus the spacetime that fit our purpose is static and a spherically symmetric. 
  Therefore the metric ansatz for the interior star is taken as,
  \begin{eqnarray}
   ds_{-}^2&=&g_{\mu\nu} dx^{\mu}dx^{\nu}\nonumber\\
   &=&-e^{\nu(r)}dt^2 + e^{\lambda(r)}dr^2 + r^2\big(d\theta^2 + \sin^2\theta d\phi^2\big),
   \label{4D metric ansatz}
  \end{eqnarray}
  
  where $\nu(r)$ and $\lambda(r)$ are arbitrary functions of radial coordinate $r$ that we need 
  to determine from gravitational equations. Such spherically symmetric spacetime 
  ensures that $\Psi$ as well as $\rho$ and $p$ are the functions of $r$ only. Hence 
  eqn.(\ref{em for radion}) can be simplified and as a consequence the 
  energy-momentum tensors of matter field, radion field are given by
  \begin{eqnarray}
  T^{\mu}_{\nu}(matter) = diag\bigg(-\rho(r), p(r), p(r), p(r)\bigg),
  \label{em for matter 2}
  \end{eqnarray}
  \begin{eqnarray}
   T^{\mu}_{\nu}(\Psi) = diag\bigg(&-&f^2(r)-U_{rad}(r), f^2(r)-U_{rad}(r),\nonumber\\ 
   &-&f^2(r)-U_{rad}(r), -f^2(r)-U_{rad}(r)\bigg)
   \label{em for radion 2}
   \end{eqnarray}
   
  respectively, where $f^2(r)$ is defined as $f^2(r) = \frac{1}{2}e^{-\lambda(r)}\Psi'(r)^2$ 
  and $U_{rad}(r) = U_{rad}(\Psi(r))$. Considering the interior of the stellar object 
  to be filled with perfect fluid having energy-momentum tensor presented in eqn.(\ref{em for matter 2}), the gravitational 
  equations (for the metric ansatz mentioned in eqn.(\ref{4D metric ansatz})) become,
  \begin{eqnarray}
   e^{-\lambda}\bigg[\frac{1}{r^2} - \frac{\lambda'}{r}\bigg] - \frac{1}{r^2} = 8\pi G_{4}\bigg[-\rho - f^2 - U_{rad}(r)\bigg],
   \label{eqn1}
  \end{eqnarray}
  \begin{eqnarray}
   e^{-\lambda}\bigg[\frac{\nu'}{r} + \frac{1}{r^2}\bigg] - \frac{1}{r^2} = 8\pi G_{4}\bigg[p + f^2 - U_{rad}(r)\bigg],
   \label{eqn2}
  \end{eqnarray}
  
  where $'$ denotes the derivative with respect to $r$ and $\frac{1}{8\pi G_4} = \frac{M^3}{k} \sim 10^{38}$(GeV)$^2$. 
  There exists another Einstein's equation corresponding to angular coordinate, but that can be derived from the above two 
  and hence is not independent.\\
  
  On the other hand, the conservation equation for the fluid and the field equation for radion field takes the 
  following simple form in the context of spherically symmetric spacetime,
  \begin{eqnarray}
   p' + \frac{\nu'}{2}\big(\rho + p\big) = 0
   \label{eqn3}
  \end{eqnarray}
  and
  \begin{eqnarray}
   f' + \frac{\nu'}{2}f + \frac{2}{r}f = \frac{U_{rad}'}{2f}
   \label{eqn4}
  \end{eqnarray}
  respectively. To derive the radion field equation, we use the definition of $f^2(r)$ as mentioned earlier. 
  At this stage, it deserves mentioning that there are four independent differential equations governing the 
  behaviour of the system considered in the present case, while there are five unknowns, $\lambda(r)$, 
  $\nu(r)$, $\rho(r)$, $p(r)$, $f(r)$. This problem is generally resolved by assuming an equation of state 
  for the perfect fluid. However this equation of state is not needed in the present context, 
  because here we are interested on the upper bound 
  of the mass-radius ratio of the star (in presence of modulus field), for which the complete interior solutions 
  are not necessary\\
  
  Next we try to get some information about the functions $\lambda(r)$, $\nu(r)$ from the above equations of motion. 
  It is easy to show that one can integrate eqn. (\ref{eqn1}), resulting into the following form of $\lambda(r)$,
  \begin{eqnarray}
   e^{-\lambda(r)}&=&1 - \frac{2G_4}{r} \int_{0}^{r} 4\pi r^2 \bigg[\rho + f^2 + U_{rad}(r)\bigg]\nonumber\\
   &=&1 - \frac{2G_4 m(r)}{r},
   \label{information of lambda}
  \end{eqnarray}
  
  where $m(r) = \int_{0}^{r} 4\pi r^2 \bigg[\rho + f^2 + U_{rad}(r)\bigg]$, the mass of the star up to radius $r$. 
  It is clear from the expression of $m(r)$ that due to the presence of radion field, the total gravitational 
  mass is different from that the actual matter density present inside. The extra gravitating mass comes from 
  the radion field strength and its potential.\\
  
  However, eqn. (\ref{eqn3}) can be rewritten as,
  \begin{eqnarray}
   \nu'(r) = \frac{-2p'}{\big(\rho + p\big)}
   \label{information of nu}
  \end{eqnarray}
  which is the famous Tolman-Oppenheimer-Volkoff (TOV) equation. 
  The matter density ($\rho$) as well as the pressure ($p$) inside the star generally decreases with an increase of $r$. 
  This behaviour of $p$ along with eqn. (\ref{information of nu}) 
  indicate that the function $\nu(r)$ increases with $r$ (i.e $\nu'(r)>0$). Moreover eqn.(\ref{eqn4}) can be 
  integrated and has a solution of $f(r)$ given by: $f^2(r) = \frac{1}{r^4}e^{-\nu} + U_{rad}(r)$. Considering $U_{rad}' < 0$ and
  due to the fact $\nu'(r) > 0$, the above expression of $f(r)$ clearly implies that $f(r)$ decreases with the radial coordinate $r$. 
  Hence the effective density (inside the star) $\rho + f^2 + U_{rad}$ also decreases as the surface of the star is approached. We 
  will use this result later on.\\
  
  With these ingredients, let us now derive Buchdahl's limit explicitly and for that let us start by 
  differentiating both sides of eqn.(\ref{eqn2}) (with respect to $r$) and get,
  \begin{eqnarray}
   e^{-\lambda}\bigg[\frac{\nu''}{r} - \frac{\nu'}{r^2} - \frac{2}{r^3} - \frac{\lambda'\nu'}{r}&-&\frac{\lambda'}{r^2}\bigg] 
   + \frac{2}{r^3} = 8\pi G_4\nonumber\\ 
   &\bigg[&p' + 2ff' - U_{rad}'(r)\bigg].
   \label{diff 1}
  \end{eqnarray}
  
  Using the conservation equations for the fluid and the radion field from eqn.(\ref{eqn3}) and eqn.(\ref{eqn4}), one can 
  evaluate the right hand side of eqn.(\ref{diff 1}), leading to
  \begin{eqnarray}
   8\pi G_4 \bigg[p' + 2ff' - U_{rad}'(r)\bigg]&=&8\pi G_4 \bigg[-\frac{\nu'}{2}\big(\rho + p\big) 
   + 2f\big(-\frac{\nu'}{2}f - \frac{2}{r}f\big)\bigg]\nonumber\\
   &=&e^{-\lambda}\bigg[-\frac{\nu'^2}{2r} - \frac{\nu'\lambda'}{2r}\bigg] - \frac{32\pi G_4}{r}f^2
   \label{rhs}
  \end{eqnarray}
  
  Substituting the above expression back to eqn.(\ref{diff 1}) and a little simplification leads to the following equation,
  \begin{eqnarray}
   2r\nu'' + r\nu'^2 - r\nu'\lambda' - 2\nu' = \frac{4}{r}\big(1 - e^{\lambda}\big) + 2\lambda' - 64\pi G_4 rf^2e^{\lambda}.
   \label{simplification 1}
  \end{eqnarray}
  
  By using the following two identities namely 
  \begin{eqnarray}
   \frac{d}{dr}\bigg[\frac{1}{r}e^{-\lambda/2}\frac{de^{\nu/2}}{dr}\bigg]&=&\frac{e^{(\nu-\lambda)/2}}{4r^2}
   \big[2r\nu'' + r\nu'^2 - r\nu'\lambda' - 2\nu'\big]
   \label{identity 1}\\
   \frac{d}{dr}\bigg[\frac{1 - e^{-\lambda}}{2r^2}\bigg]&=&\frac{e^{-\lambda}}{2r^3} \big[r\lambda' - 2\big(e^{\lambda} - 1\big)\big]
   \label{identity 2}
  \end{eqnarray}
  
  Eqn.(\ref{simplification 1}) can be rewritten as follows,
  \begin{eqnarray}
   e^{-(\nu+\lambda)/2} \frac{d}{dr}\bigg[\frac{1}{r}e^{-\lambda/2}\frac{de^{\nu/2}}{dr}\bigg] = 
   \frac{d}{dr}\bigg[\frac{1 - e^{-\lambda}}{2r^2}\bigg] - \frac{16\pi G_4}{r}f^2
   \label{simplification 2}
  \end{eqnarray}
  
  At this point, we put forward some sensible requirements: the average energy density inside the star $\rho_{av} = m(r)/r^3$ should decrease 
  with the radial coordinate. Even though the average density involves contribution from the radion field, since 
  the radion field strength itself decreases outwards, the above requirement will trivially hold. Further, the form of 
  $e^{-\lambda}$ given in eqn.(\ref{information of lambda}) indicates that the first term of the right hand side of 
  eqn.(\ref{simplification 2}) is essentially $d\rho_{av}/dr$ i.e. $\frac{d}{dr}\bigg[\frac{1 - e^{-\lambda}}{2r^2}\bigg] = \frac{d\rho_{av}}{dr}$. 
  This expression along with the decreasing character of $\rho_{av}$ (with $r$) makes the right hand side of eqn.(\ref{simplification 2}) 
  negative. As a consequence, we get the following inequality:
  \begin{eqnarray}
   \frac{d}{dr}\bigg[\frac{1}{r}e^{-\lambda/2}\frac{de^{\nu/2}}{dr}\bigg] < 0.
   \label{inequality 1}
  \end{eqnarray}
  
  Integrating the above relation from some radius $r$ within the star to the surface of the star, 
  given by the radius $r_0$, we obtain,
  \begin{eqnarray}
   \frac{de^{\nu/2}}{dr} > \frac{\nu_0'}{2r_0} re^{\lambda/2} e^{(\nu_0 - \lambda_0)/2},
   \label{inequality 2}
  \end{eqnarray}
  
  where the quantities with the subscript $'0'$ denotes that they are to be evaluated at the surface of the star 
  i.e. at $r = r_0$. Furthermore, to derive the above inequality, we consider that both the metric and its first derivative 
  are continuous at $r = r_0$. However, later, in section [4], 
  we explicitly find the continuity conditions of metric and its first derivative at the boundary of 
  the star by considering a generalized Vaidya metric for the exterior spacetime of the stellar object.\\
  
  Integrating again the both sides of eqn.(\ref{inequality 2}) from the origin to the surface of the star, we obtain
  \begin{eqnarray}
   e^{\nu_0/2} - e^{\nu/2}(r=0)&>&\frac{\nu_0'}{2r_0} e^{(\nu_0 - \lambda_0)/2}\int_0^{r_0} re^{\lambda/2} dr\nonumber\\
   &=&\frac{\nu_0'}{2r_0} e^{(\nu_0 - \lambda_0)/2}\int_0^{r_0} dr \frac{r}{\sqrt{1 - \frac{2G_4 m(r)}{r}}}.
   \label{inequality 3}
  \end{eqnarray}
  In the last line, we use the solution of $e^{\lambda}$ (see eqn.(\ref{information of lambda})). 
  As the average energy density decreases towards the boundary of the star, it immediately follows that 
  $m(r)/r^3 > M/r^3$, where $M = m(r_0)$, the total effective mass of the star. Thus the inequality 
  in eqn.(\ref{inequality 3}) holds more strongly if $\frac{m(r)}{r}$ is replaced by $\frac{M}{r_0^3}r^2$. 
  With this modification, we arrive at,
  \begin{eqnarray}
   e^{\nu_0/2} - e^{\nu/2}(r=0)&>&\frac{\nu_0'}{2r_0} e^{(\nu_0 - \lambda_0)/2}\int_0^{r_0} dr \frac{r}{\sqrt{1 - \frac{2G_4M}{r_0^3}r^2}}\nonumber\\
   &=&\frac{\nu_0'}{2r_0} e^{(\nu_0 - \lambda_0)/2} \bigg(\frac{r_0^3}{2G_4M}\bigg) \bigg[1 - \sqrt{1- \frac{2G_4M}{r_0}}\bigg].
   \label{inequality 4}
  \end{eqnarray}
  
  Both the pressure and the contribution of the radion field are positive and finite at the origin, it 
  follows that $e^{\nu/2}(r=0) > 0$. Applying this result into eqn.(\ref{inequality 4}), we immediately obtain the 
  following inequality,
  \begin{eqnarray}
   1 - \frac{\nu_0'}{2r_0} e^{-\lambda_0/2} \bigg(\frac{r_0^3}{2G_4M}\bigg) \bigg[1 - \sqrt{1- \frac{2G_4M}{r_0}}\bigg] > 0.
   \label{inequality 5}
  \end{eqnarray}
  
  The factor $\nu_0'$ can be obtained in terms of $\lambda_0$ by considering eqn.(\ref{eqn2}) at the surface 
  of the star ($r=r_0$, where the pressure is zero i.e. $p(r_0)=0$) as,
  \begin{eqnarray}
   \frac{\nu_0'}{2r_0} = -\frac{1}{2r_0^2} + \frac{1}{2}e^{\lambda_0} \bigg[\frac{1}{r_0^2} + 8\pi G_4\big(f_0^2 - U_{rad}^0\big)\bigg].
   \label{intermediate}
  \end{eqnarray}
  Recall $f_0^2 = \frac{1}{2}e^{-\lambda_0}\Psi'(r_0)^2$ and $U_{rad}^0$ is the radion potential at $r_0$. Plugging the 
  above expression of $\nu_0'$ into eqn.(\ref{inequality 5}) and further using the solution of $e^{\lambda}$, we finally lands up with the 
  inequality as follows:
  \begin{eqnarray}
   \sqrt{1 - \frac{2G_4M}{r_0}}&-&\frac{r_0^3}{2G_4M}\bigg(1 - \sqrt{1 - \frac{2G_4M}{r_0}}\bigg)\nonumber\\
    &\bigg[&\frac{1}{2r_0^2}\bigg(\frac{2G_4M}{r_0}\bigg) + 4\pi G_4\big(f_0^2 - U_{rad}^0\big)\bigg] > 0.
    \label{inequality 6}
  \end{eqnarray}
  
  Simplification of eqn.(\ref{inequality 6}) yields a quadratic expression for $\frac{2G_4M}{r_0}$,  one root of which  
  corresponds to a negative value and hence can be safely ignored, while the other root provides the necessary limit 
  on mass-radius ratio (i.e. $\frac{M}{r_0}$) of the star, as follows,
  \begin{eqnarray}
   \frac{2G_4M}{r_0} < \frac{b}{2a} \bigg[1 + \sqrt{1 + \frac{4ac}{b^2}}\bigg]
   \label{final inequality}
  \end{eqnarray}
  
  where $a$, $b$, $c$ have the following expressions:
  \begin{eqnarray}
   a = \frac{9}{4} - 6\pi G_4r_0^2\Psi_0'^2 + 4\pi^2G_4^2r_0^4\Psi_0'^4,
   \label{a}
   \end{eqnarray}
   \begin{eqnarray}
   b = 2 - 10\pi G_4r_0^2\Psi_0'^2 + 8\pi^2G_4^2r_0^4\Psi_0'^4 + 12\pi G_4r_0^2U_{rad}^0 - 16\pi^2G_4^2r_0^4\Psi_0'^2U_{rad}^0,
   \label{b}
   \end{eqnarray}
   \begin{eqnarray}
   c&=&4\pi G_4r_0^2\Psi_0'^2 - 4\pi^2G_4^2r_0^4\Psi_0'^4 - 8\pi G_4r_0^2U_{rad}^0 - 16\pi^2G_4^2r_0^4(U_{rad}^0)^2\nonumber\\
   &+&16\pi^2G_4^2r_0^4\Psi_0'^2U_{rad}^0.
   \label{c}
  \end{eqnarray}
  
  Eqns.(\ref{a}) to (\ref{c}) indicate that in the absence of the radion field, $a$, $b$, $c$ take the value as 
  $\frac{9}{4}$, $2$, $0$ respectively, for which one immediately recovers the usual Buchdahl's limit in General Relativity, 
  i.e $\frac{2G_4M}{r_0} < \frac{8}{9}$.\\
  
\begin{figure}[!h]
\begin{center}
 \centering
 \includegraphics[width=4.2in,height=4.2in]{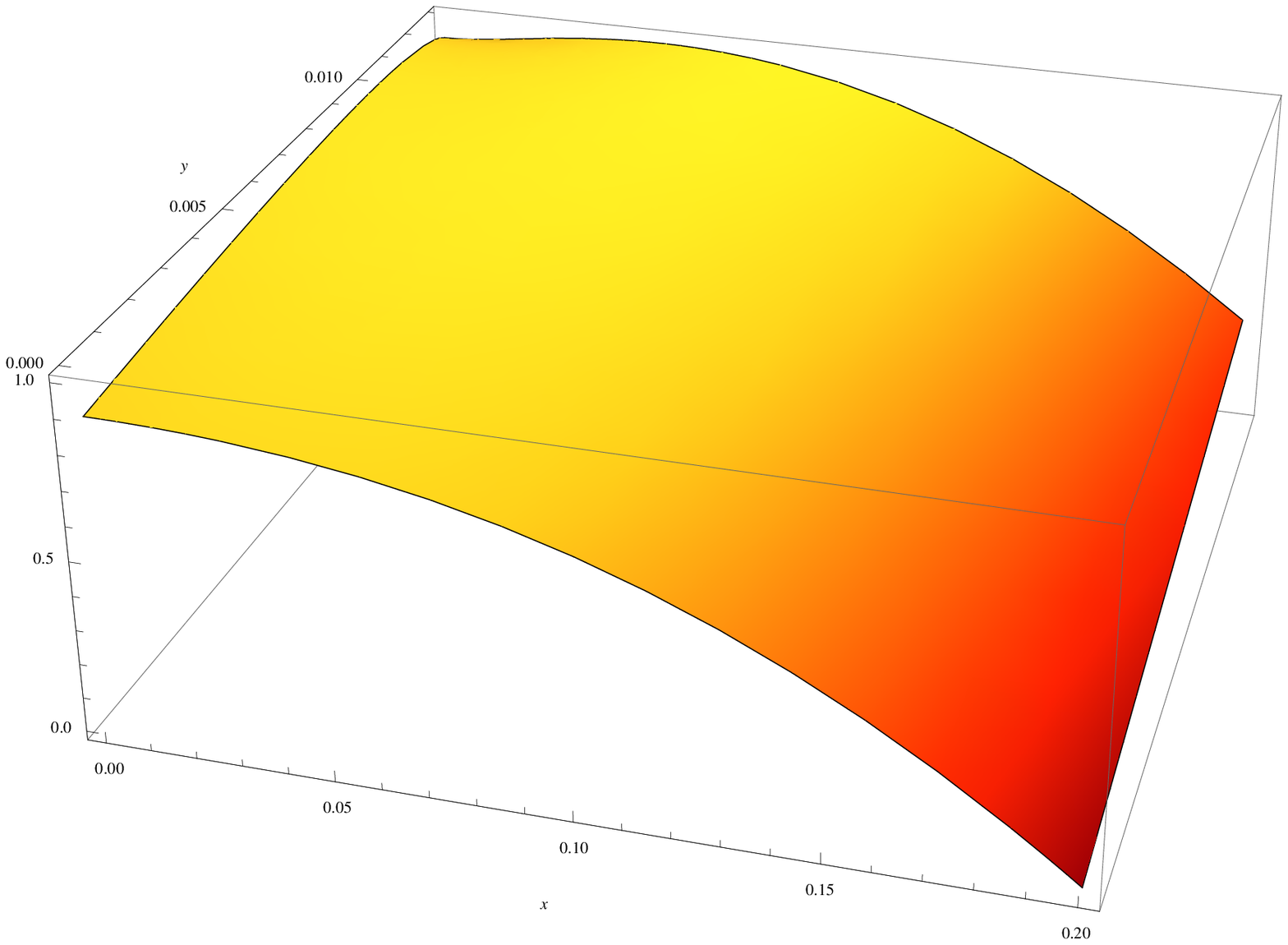}
 \includegraphics[width=0.3in,height=2.0in]{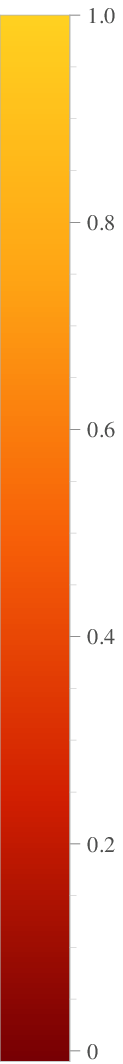}
 \caption{$\frac{2G_4M}{r_0}$ vs $G_4r_0^2\Psi_0'^2 (=x)$ and $G_4r_0^2U_{rad}^0 (=y)$ (left figure), corresponding contour plot (right figure)}
 \label{plot stellar limit}
\end{center}
\end{figure}

  However, in the presence of the radion field ($\Psi$), the upper limit on mass-radius ratio of a stable stellar object 
  gets modified compared to general relativity and obviously depends on the strength of the radion field 
  (and its potential) on the surface of the star. From eqn.(\ref{final inequality}) (along with the expressions of $a$, $b$, $c$), we obtain 
  figure (\ref{plot stellar limit}) demonstrating the variation of $\frac{2G_4M}{r_0}$ with $f_0$ and $U_{rad}^0$.\\
  
Figure (\ref{plot stellar limit}) (with the contour plot) clearly reveals that in the presence of higher 
dimensional modulus field, the upper bound on $\frac{2G_4M}{r_0}$ can be larger than $8/9$ 
and reach up to unity. Therefore extra mass 
can be packed into the stellar structure in comparison to the Einstein 
gravity. This provides an interesting testbed for existence of the modulus field (or radion field) 
which carries the footprint of compactified extra dimension on our visible universe. 
Thus, if a compact stellar object (say a neutron star) 
is observed whose $2G_4M/r_0$ is larger than $8/9$ $\big($i.e $\frac{2G_4M}{r_0}$ lies between [8/9,1]$\big)$, 
then it can possibly signal towards the existence of such higher dimension.\\

\section{Matching of the interior spacetime with an exterior geometry}

To complete the model, the interior spacetime geometry of the spherical star needs to be matched to 
an exterior geometry. For the required matching, the Israel conditions are used, 
where the metric coefficients and extrinsic curvatures (first and second fundamental forms 
respectively) are matched at the boundary of the sphere.\\ 

However in the presence of a scalar field (which is the radion field in the present context), the interior 
spacetime can not be smoothly matched to a vacuum exterior (i.e. the Schwarzschild one). If the exterior is a vacuum, the 
scalar field has to behave as a delta function at the boundary resulting in a square of delta function for the energy density.\\

To avoid this problem, in the present case, we match 
the interior spacetime with a generalized Vaidya exterior spacetime at the boundary hypersurface $\Sigma$ 
given by $r = r_0$. The metric inside and outside of $\Sigma$ are given by,
\begin{eqnarray}
 ds_{-}^2 = -e^{\nu(r)}dt^2 + e^{\lambda(r)}dr^2 + r^2\big(d\theta^2 + \sin^2\theta d\phi^2\big)
 \label{inside metric}
\end{eqnarray}
and
\begin{eqnarray}
 ds_{+}^2 = -\bigg(1 - \frac{2G_4M_+(r_v)}{r_v}\bigg)dv^2 + 2dvdr_v + r_v^2\big(d\theta^2 + \sin^2{\theta}d\varphi^2\big)
 \label{outside metric}
\end{eqnarray}
respectively, where $r_v,v,\theta$ and $\varphi$ are the exterior coordinates and $M_+(r_v)$ (the suffix $'+'$ stands for exterior) 
is exterior mass function, which is independent of $v$ due to the reason that the spacetime is static. 
The same hypersurface $\Sigma$ can alternatively be defined by the exterior 
coordinates as $r_v = R(t)$ and $v = T(t)$. Then the metrics on $\Sigma$ from inside and 
outside coordinates turn out to be,
\begin{eqnarray}
 ds_{-,\Sigma}^2 = -e^{\nu_0}dt^2 + r_0^2 d\Omega^2
 \nonumber
\end{eqnarray}
and
\begin{eqnarray}
 ds_{+,\Sigma}^2 = -\bigg[\bigg(1 - \frac{2G_4M_{+,\Sigma}}{R(t)}\bigg)\dot{T}^2 - 2\dot{T}\dot{R}\bigg]dt^2 + R(t)^2 d\Omega^2,
 \nonumber
\end{eqnarray}
where $M_{+,\Sigma}$ is the exterior mass function on $\Sigma$, $d\Omega^2$ denotes the line 
element on a unit two sphere and dot represents $\frac{d}{dt}$. Matching the first fundamental 
form on $\Sigma$ (i.e. $ds_{-,\Sigma}^2 = ds_{+,\Sigma}^2$) yields the following two conditions :
\begin{eqnarray}
 \frac{dT(t)}{dt} = \frac{e^{\nu_0/2}}{\sqrt{1 - \frac{2G_4M_{+,\Sigma}}{r_0}}}
 \label{con 1}
\end{eqnarray}
and
\begin{eqnarray}
 R(t) = r_0.
 \label{con 2}
 \end{eqnarray}
In order to match the second fundamental form, we calculate the normal of the hypersurface $\Sigma$ 
from inside ($\vec{n}_{-} = n_{-}^t$, $n_{-}^r$, $n_{-}^{\theta}$, $n_{-}^{\varphi}$) and outside 
($\vec{n}_{+} = n_{+}^v$, $n_{+}^{r_v}$, $n_{+}^{\theta}$, $n_{+}^{\varphi}$) coordinates as follows,
\begin{eqnarray}
 n_{-}^t = 0~~,~~~~~~~~n_{-}^r =  e^{-\lambda_0/2}~~,~~~~~~~n_{-}^{\theta} = n_{-}^{\varphi} = 0\label{inside normal}
\end{eqnarray}
and
\begin{eqnarray}
 n_{+}^v = \frac{1}{\sqrt{1 - \frac{2G_4M_{+,\Sigma}}{r_0}}}~~,\nonumber
 \end{eqnarray}
 \begin{eqnarray}
 n_{+}^{r_v} = \sqrt{1 - \frac{2G_4M_{+,\Sigma}}{r_0}}~~,\nonumber
 \end{eqnarray}
 \begin{eqnarray}
 n_{+}^{\theta} = n_{+}^{\varphi} = 0.
\label{outside normal}
\end{eqnarray}

To derive the normal vectors, we use eqn.(\ref{con 2}). 
The above expressions of $\vec{n}_{-}$ and $\vec{n}_{+}$ leads to the extrinsic 
curvature of $\Sigma$ from interior and exterior coordinates respectively, and are given by,
\begin{eqnarray}
 K_{tt}^- = -\frac{\nu_0'}{2}e^{\nu_0}e^{-\lambda_0/2}~~,\nonumber
 \end{eqnarray}
 \begin{eqnarray}
 K_{\theta\theta}^- = r_0 e^{-\lambda_0/2}~~,~~~~~~~K_{\varphi\varphi}^- = r_0 e^{-\lambda_0/2}\sin^2{\theta},
 \label{inside extrinsic}
\end{eqnarray}
from interior metric, and
\begin{eqnarray}
 K_{tt}^+ = -G_4 \frac{e^{\nu_0}}{\sqrt{1 - 2\frac{G_4M_{+,\Sigma}}{r_0}}} \bigg[\frac{M_{+,\Sigma}}{r_0^2} 
 - \frac{1}{r_0}\frac{\partial M_{+}(r_v)}{\partial r_v}\bigg{|}_{\Sigma}\bigg]~~,
 \nonumber
\end{eqnarray}
\begin{eqnarray}
 K_{\theta\theta}^+ =  r_0 \sqrt{1 - \frac{2G_4M_{+,\Sigma}}{r_0}}~~,
 \nonumber
\end{eqnarray}
\begin{eqnarray}
 K_{\varphi\varphi}^+ = r_0 \sqrt{1 - \frac{2G_4M_{+,\Sigma}}{r_0}} \sin{\theta}^2
 \label{outside extrinsic}
\end{eqnarray}
from exterior metric.\\

The equality of the extrinsic curvatures of $\Sigma$ from both sides is therefore equivalent to the following 
two conditions :
\begin{eqnarray}
 e^{-\lambda_0)/2} = \sqrt{1 - \frac{2G_4M_{+,\Sigma}}{r_0}}
 \label{con 3}
\end{eqnarray}
and
\begin{eqnarray}
 \frac{\partial M_{+}(r_v)}{\partial r_v}\bigg{|}_{\Sigma} = \frac{M_{+,\Sigma}}{r_0} 
 - \frac{r_0}{2G_4}\nu_0' \bigg[1 - \frac{2G_4M_{+,\Sigma}}{r_0}\bigg].
 \label{con 4}
\end{eqnarray}

The total mass of the spherical star is given by: $M = \int_0^{r_0} 4\pi r^2 \big(\rho + f^2 + U_{rad}(r)\big)$. 
Eqn.(\ref{con 3}) along with the solution of $\lambda$ (see eqn.(\ref{information of lambda})) relates the exterior 
mass function on $\Sigma$ (i.e. $M_{+,\Sigma}$) with the total mass of the stellar object as,
\begin{eqnarray}
 M_{+,\Sigma}&=&M\nonumber\\
 &=&\int_0^{r_0} 4\pi r^2 \big(\rho + f^2 + U_{rad}(r)\big).
 \label{con 5}
\end{eqnarray}

Eqns. (\ref{con 1}), (\ref{con 2}), (\ref{con 4}), (\ref{con 5}) completely 
specify the matching at the boundary of the star with an exterior generalized Vaidya geometry.\\

\section{Conclusion}
We consider a five dimensional AdS compactified warped geometry model with two 3-branes embedded within the spacetime. 
For the purpose of modulus stabilization, a massive scalar field is invoked in the bulk and its backreaction 
on spacetime geometry is taken into account. In such a scenario, our universe is identified with a 3-brane and emerges 
as a four dimensional effective theory. On projecting the bulk gravity on the brane, the extra degrees of freedom of 
$R^{(5)}$ appears as a scalar field (known as radion field) in the 4D effective on-brane theory. 
From the perspective of such on-brane theory, we explore the effect of radion field on the limit 
of mass-radius ratio ($M/r_0$) for a stable stellar structure. \\

We match the interior spacetime of the star with a suitable exterior geometry on the boundary ($\Sigma$). For this 
 matching, the Israel junction conditions are used where the metric coefficients and extrinsic curvatures are matched on $\Sigma$. 
 At this stage, it deserves mention that in presence of a scalar field (which is the radion field in the present context), 
 the matching of interior spacetime with exterior Schwarzschild geometry leads to some inconsistency. For instance, since 
 Schwarzschild has zero scalar field, such a matching would lead to a discontinuity 
 in the scalar field, which means a delta function in the gradient of the scalar field. 
 As a consequence, there will appear square of a delta function in the stress-energy, which is definitely an inconsistency. 
 To avoid such problems, here we consider the exterior geometry as a generalized Vaidya spacetime. With this consideration, 
 we determine the matching conditions given in eqns. (\ref{con 1}), (\ref{con 2}), (\ref{con 4}), (\ref{con 5}). \\

 The main conclusion of the present investigation is the following. Due to the presence of radion field, the upper limit on mass-radius ratio (generally known as Buchdahl's limit) of 
 a compact stellar object gets modified in comparison to general relativity and obviously depends on the 
 strength of the radion field (and its potential) on the surface of the star. The variation of Buchdahl's limit 
 with the radion field strength is shown in figure (\ref{plot stellar limit}), which 
 clearly demonstrates that in the presence of the higher dimensional modulus field, the upper bound of $\frac{2G_4M}{r_0}$ can go 
 beyond the value $8/9$ and reach up to unity; while the general relativity prediction is given by: $\frac{2G_4M}{r_0} < \frac{8}{9}$. 
 Therefore extra mass can be packed into the stellar structure in comparison to the Einstein gravity. 
 This provides an interesting testbed for existence of modulus field (or radion field) 
 which carries the footprint of compactified extra dimension on our visible universe. Hence if it is possible to detect 
 a compact object with mass-radius ratio larger than the general relativity prediction, then one can infer about 
 the possible presence of such higher dimension.\\

 \section*{Acknowledgements}
  The author would like to thank Narayan Banerjee and Soumitra SenGupta for illuminating discussions.

\end{document}